\documentstyle[amsfonts,epsfig,12pt]{article}
\newcommand{\be}{\begin{equation}}
\newcommand{\ee}{\end{equation}}
\newcommand{\bea}{\begin{eqnarray}}
\newcommand{\eea}{\end{eqnarray}}
\newcommand{\beas}{\begin{eqnarray*}}
\newcommand{\eeas}{\end{eqnarray*}}
\newcommand{\bi}{\begin{itemize}}
\newcommand{\ei}{\end{itemize}}
\newcommand{\bc}{\begin{center}}
\newcommand{\ec}{\end{center}}
\newcommand{\bfl}{\begin{flushleft}}
\newcommand{\efl}{\end{flushleft}}
\newcommand{\bfr}{\begin{flushright}}
\newcommand{\efr}{\end{flushright}}

\def\6{\partial} \def\a{\alpha} \def\b{\beta}
\def\g{\gamma}

  \def\l{\lambda}
\def\m{\mu} \def\n{\nu}  
 \def\s{\sigma} 
  
\def\o{\omega} \def\G{\Gamma}

\newcommand{\DD}{{\cal D}}

\newcommand{\HH}{{\cal H}}

\newcommand{\LL}{{\cal L}}

\newcommand{\OO}{{\cal O}}

\begin{document}
\title{Thermal D-brane boundary states from type IIB Green-Schwarz superstring 
in pp-wave background}
\author{{{Ion V. Vancea\footnote{ionvancea@ufrrj.br}}}}

\maketitle

\begin{center}
{{\em Departamento de F\'{\i}sica, Universidade Federal Rural do Rio de Janeiro},\\
{\em Cx. Postal 23851, 23890-000 Serop\'{e}dica - RJ, Brasil}}
\end{center}

\begin{abstract}
We construct the thermal boundary states from the type IIB Green-Schwarz superstring in {\em pp}-wave background in the light-cone gauge. The superstring is treated in the canonical ensemble and in the TFD formalism which is appropriate to discuss canonically quantized systems. The thermal boundary states are obtained by thermalizing the total boundary states which are the boundary states of the total system that is composed by the superstring modes and the corresponding thermal reservoir modes. That analysis is similar to the one in the flat spacetime case \cite{ivv12}. However, there are some subtleties concerning the construction of the total string which are discussed. Next, we compute the entropy of thermal boundary state which is defined as the expectation value of the superstring entropy operator in the thermal boundary state.  
\end{abstract}

\newpage

\section{Introduction}

Recently, a considerable attention has been devoted to the string theory formulated in {\em pp}-wave background. The interest has been triggered by the remarkable proof of the exact perturbative quantization of the type IIB Green-Schwarz (GS) superstring in the light-cone gauge in the presence of the {\em pp}-waves \cite{rrm,mt} and of the relationship between the string theory in {\em pp}-wave background and the {\em AdS/CFT} correspondence in the Penrose limit of the $AdS \times S$. The non-triviality of that background and the fact that the string and brane dynamics in the presence of the {\em pp}-waves can be investigated using a large variety of techniques, make the {\em pp}-wave background particularly interesting for studying the string and brane thermodynamics. 

One reason for which it is important to analyse the thermodynamical properties of strings and branes in general backgrounds is that from string theory point of view, the string thermodynamics is crucial for understanding the early Universe, the cosmology and the thermodynamics of black-holes, all of which, according to our present knowledge, should display non-trivial backgrounds. Another fundamental motivation is that the Statistical Mechanics and the Thermodynamics of the extended objects is not a well understood subject, yet, mainly due to the problems related to the presence of the gravity in the theory. Therefore, the investigation of the thermal properties of strings and branes in tractable backgrounds, such as the {\em pp}-wave background, could help to improve our understanding of the general theory of string and brane thermodynamics.

Most of the thermal string properties can be analysed using the general concepts of the Statistical Mechanics only in the perturbative limit of the string theory in the flat spacetime \cite{aw}. But whenever the canonical quantization methods are available, these concepts can be extended, at least formally, to other backgrounds, as in the case of the {\em pp}-wave background. Indeed, progress has been made in the study of the thermal properties of the type IIB GS superstring in the {\em pp}-wave background with a focus on the construction of the string thermal partition function and free energy and the derivation of the Hagedorn temperature \cite{zv,gss,ys1,blt,ys2,gost,bc}. The free energy of the type IIA GS superstring was studied in \cite{hpy}. The thermal instabilities of the massless string modes and of the membranes at finite temperature on {\em pp}-waves were discussed in the context of the matrix model theory in \cite{whh,sy}. On a different research line, it has been proved that, as in the flat background, there is a relationship among the perturbative closed superstring modes in the light-cone gauge on a {\em pp}-wave and the $D$-branes defined by the Dirichlet boundary conditions on the cylinder-dual open superstring. The corresponding closed superstring coherent states localized on the Dirichlet hypersurfaces were calculated for the type IIB GS superstrings in \cite{bp,dp} following the same hypothesis as in the flat spacetime \cite{gg1} (see also \cite{st1,st2,st3}). The calculation of their normalization constant was given in \cite{bgg}. Subsequently, the general properties of the type IIB boundary states were analysed in \cite{bp,dp,bgg,hy,gg,st,ggss} and that of the type IIA boundary states were studied in \cite{kp,cly1,dr,cly2,mg,ak,tm1,gks,llp,tm2,pm} . Thus, the existence of the thermal string, on one hand, and of the coherent state representation of $D$-branes, on the other hand, suggests the existence of a relationship between the thermalized string states and the thermalized $D$-brane boundary states in the {\em pp}-wave background analogous to the one in the flat spacetime \cite{ivv1}. The aim of the present paper is to construct the thermal $D$-brane boundary states from the thermal closed superstring boundary states obtained by thermalization of the closed superstring at zero temperature in {\em pp}-wave background. 

The study of the thermal boundary states is conveniently performed by applying the thermalization method based on the Thermo Field Dynamics (TFD) \cite{ubook} and developed in \cite{ivv1,ivv2,ivv3,ivv4,ivv5,ivv6}. According to this method, the superstring modes in the canonical ensemble interact with modes of equal frequency from the thermal reservoir. Thus, the reservoir can be represented by a copy of the initial superstring. The thermal interaction among the superstring modes and the identical reservoir modes is represented by the Bogoliubov operator constructed from the creation and annihilation operators of all modes. As a result of this interaction, the original superstring heats up to some finite temperature. However, the boundary conditions should not be affected by the heating since they represent the geometrical localization of the world-sheet boundaries in spacetime, which should not change when the temperature varies smoothly from zero to some finite value. The simplest way in which these heuristic ideas can be formally implemented in superstring theory is by constructing a unitary TFD operator which maps the superstring modes as well as the boundary conditions from $T=0$ to some $T \neq 0$ (as in the references cited above). That amounts to working within the canonical quantization formalism. That is particularly convenient in the present case because the canonical formalism is better suited for analysing the coherent boundary states in the superstring theory. The TFD was applied previously to the string theory in \cite{yl1,yl2,yl3,yl4,yl5,yl6,fns1,fn1,fn2,fn3,fn4,fn5,f1}. More recently, various aspects of the thermal string in the TFD formulation were discussed in \cite{ivv7,ivv8,ivv9,ivv10,ng3,ng4,ng5,mb}. Of particular interest for this work are the references \cite{ng1,ng2} where the TFD formalism was applied to the GS superstring and string field theory in the {\em pp}-wave background and \cite{ivv11,ivv12} where the thermal $D$-branes were constructed from the boundary states of the type IIB GS superstring in the light-cone gauge in flat spacetime. Thus, the present work represents the generalization of the method cited above for constructing thermal $D$-brane states from Minkowski spacetime to the curved {\em pp}-wave background. 

The paper is organized as folows. In Section 2 we review the $D$-branes in the type IIB GS superstring in the light-cone gauge. In Section 3 we thermalize the superstring and construct the thermal boundary states. To this end, 
we map the $D$-brane boundary conditions from $T = 0 $ to $ T \neq 0$ by applying the unitary thermal operator obtained by exponentiating the Bogoliubov operator. Then we show that the solution to the thermal boundary conditions can be obtained from the zero temperature boundary states by thermalization as in the flat spacetime. The $D$-brane entropy is calculated in Section 4. The last section is devoted to discussions and conclussions.

\section{GS superstring and D-branes in {\em pp}-wave background}

In this section we are going to review the type IIB GS superstring in the {\em pp}-wave background and the $D$-brane construction following \cite{bp,dp,bgg}. The $d=10$ {\em pp}-wave background is defined by the following
metric 
\bea
ds^{2} &=& 2dx^{+}dx^{-}-f^{2}x^{I}x^{I}dx^{+}dx^{-}+dx^{I}dx^{I},
\label{pp-metric}\\
f &=& \frac{1}{2}F_{+1234}=\frac{1}{2}F_{+5678},
\label{five-form}
\eea
where $x^{\pm} = \frac{1}{2}\left( x^9 \pm x^0 \right)$, $I,J = 1, 2, \ldots 8$ and $F^{(5)}$ is the constant Ramond-Ramond five-form field strength. The background (\ref{pp-metric}) has the maximal number of 32 supersymmetries as the flat spacetime and the $AdS_5 \times S^5$. However, the transverse isometry group $SO(8)$ is broken to $SO(4) \times SO(4)$ by the $F^{(5)}$ constant flux. As a consequence, the directions $I=1,2,3,4$ are distinct from $I=5,6,7,8$. The world-sheet string theory in the light-cone gauge is not conformally invariant in this background
since the string fields are massive. The light-cone gauge is obtained by imposing the following condition
\be
x^{+} = 2 \pi \alpha ' p^+ \tau.
\label{light-cone-gauge}
\ee
The type IIB light-cone superstring theory in the {\em pp}-wave background has eight massive scalar fields and eight massive fermionic fields $x^I, \theta^{aA}$, where $a,b = 1,2,\ldots,8$ are $SO(9,1)$ spinor indices after the {\em k}-symmetry is fixed and $A=1,2$ are world-sheet indices. The string tension can be absorbed into the fields by the following rescaling: $x^- \rightarrow 2\pi \alpha ' x^-$, $x^I \rightarrow (2\pi \alpha ')^{1/2} x^I$, $\theta^A \rightarrow (2\pi \alpha ')^{1/2} \theta^{A}$. Next one can redefine the spinors 
\be
\G^{+-}\theta^{1} = 2^{-3/4}S~~,~~\G^{+-}\theta^{2} = 2^{-3/4}\overline{S},
\label{spinor-redefinition}
\ee 
where $\G^{+-} = \frac{1}{\sqrt{2}}\left(\G^+\G^- - \G^-\G^+ \right)$, $\G^{\m}$ are the ten-dmensional Dirac matrices in the Majorana representation with $\m , \n = 0,1,\ldots,9$ and $\thetaÂ$ are real spinors \cite{rrm}. With these notations the superstring action has the following form
\begin{equation}
I = \frac{1}{2}\int_{\Sigma }d^{2}\sigma \left( 
\partial _{+}x^{I}\partial_{\_}x^{I}-m^{2}x^{I}x^{I}+
iS^{a}\partial _{+}S^{a}+i\overline{S}^{a}\partial_{-}\overline{S}^{a}-2imS^{a}\Pi _{ab}\overline{S}^{b}
\right), 
\label{action}
\end{equation}
where $\6_{\pm} = \6_0 \pm \6_1$, $m = p^+ f$, $\Pi_{ab} = \g^1 \overline{\g}^2 \g^3 \overline{\g}^4$ and $\g^{I}$ are eight-dimensional real and symmetric Dirac matrices \cite{rrm}. The action defined by the relation (\ref{action}) is invariant up to a total derivative term under two types of on-shell supersymmetry transformations: the dynamical supersymmetry
\begin{eqnarray}
\delta x^{I} &=& \a i \epsilon \gamma ^{I}\Pi S + \b i\overline{\epsilon }\gamma^{I}\Pi \overline{S},  
\label{dyn-susy1} \\
\delta S^{a} &=& \l \partial _{-}x^{I}\left( \epsilon \gamma ^{I}\Pi \right)^{a} + 
\o m x^{I}\left( \overline{\epsilon }\gamma ^{I}\Pi \right) ^{a} ,
\label{dyn-susy2}\\
\delta \overline{S}^{a} &=& \s \partial _{-}x^{I}
\left( \overline{\epsilon }\gamma ^{I}\Pi \right) ^{a} + 
\m m x^{I}\left( \overline{\epsilon }\gamma ^{I}\Pi\right)^{a}, 
\label{dyn-susy3}
\end{eqnarray}
where for future references the numerical coefficients $\a = \b = \o = \m =1$ and $\l = \s = -1$ have been put in a general form. The kinematical supersymmetry transformations have the following expressions 
\begin{eqnarray}
\delta x^{I} &=&0,  
\label{kin-susy1} \\
\delta S^{a} &=&e^{-2m \Pi \tau }\eta ^{a} + 
e^{2m \Pi \tau }\overline{\eta }^{a}, 
\label{kin-susy2}\\
\delta \overline{S}^{a} &=&e^{-2m \Pi \tau }\eta ^{a} - 
e^{2m \Pi \tau }\overline{\eta }^{a}.
\label{kin-susy3}
\end{eqnarray}
The superstring equations of motion obtained by varying the action (\ref{action}) describe the dynamics of the eight massive scalar fields and the eight massive spinors  
\begin{eqnarray}
\partial _{+}\partial _{-}x^{I}+m^{2}x^{I} &=&0, 
\label{eq-mot-bosons1} \\
\partial _{+}S^{a}-m(\Pi \overline{S})^{a} &=&0, 
\label{eq-mot-spinors2}\\
\partial _{-}\overline{S}^{a}+m(\Pi S)^{a} &=&0.
\label{eq-mot-spinors3}
\end{eqnarray}
The equations of motion (\ref{eq-mot-bosons1})-(\ref{eq-mot-spinors3}) hold if the following boundary terms vanish
\be
\delta x^{I}\partial _{\sigma }x^{I}|_{\sigma =0,\pi } = 0 ~~,~~
\delta x^{-}\partial _{\sigma }x^{+}|_{\sigma =0,\pi } = 0 ~~,~~
\left( \partial _{\sigma }x^{-}-f^{2}x^{I}x^{I}\partial _{\sigma}x^{+}\right) |_{\sigma =0,\pi } = 0,
\label{boson-boundary-term}
\ee
in the bosonic sector and 
\be
\left( S^a \delta S^a - \overline{S} ^a \delta \overline{S} ^a \right)|_{\sigma = 0, \pi} = 0,
\label{fermion-boundary-term}
\ee
in the fermionic sector, respectively. The bosonic term (\ref{boson-boundary-term}) implies that in the transverse directions the usual Neumann boundary conditions $\6_{\s} x^I|_{\s = 0, \pi} =0$ and the Dirichlet boundary conditions $\delta x^{I} |_{\s = 0, \pi} =0$, respectively,  must be imposed at the ends of the open string, as well as the periodicity condition of the closed string $x^{I}(\s = 0) = x^{I}(\s = \pi)$. In the light-cone directions the boundary conditions for $x^{\pm}$ fields are Dirichlet independently on the boundary conditions taken in the transverse directions. By using the Virasoro conditions \cite{bp} one can see that the boundary term in the fermionic sector (\ref{fermion-boundary-term}) vanishes if the fermionic boundary conditions preserve half of the background supersymmetry. That is the same ansatze as in the flat spacetime. For the open string it has the following form: $\overline{S}^a|_{\s = 0} = S^{a}|_{\s = 0}$ and $\overline{S}^{a}|_{\s = \pi } = R^{a}_{b}{S}^{b}|_{\s = \pi }$, where $R$ is a matrix from $\mathbf{{8}_{s}}$ and  $RR^T =1$, while for the closed string the fermionic boundary conditions are given by the relations $S^a|_{\s = 0} = \eta_1 S^{a}|_{\s = \pi}$ and $\overline{S}^a|_{\s = 0} = \eta_2 \overline{S}^{a}|_{\s = \pi}$ where $\eta_{1,2} = \pm 1$ \cite{bp,dp,bgg}.
The equations of motion (\ref{eq-mot-bosons1})-(\ref{eq-mot-spinors3}) with the above boundary conditions can be solved exactly \cite{rrm,bp,dp}. The Fourier expansions of the bosonic general solutions for the open and closed string have the following forms
\begin{eqnarray}
x_{NN}^{I}(\tau ,\sigma ) &=&\cos (m\tau )x_{0}^{I}+\frac{1}{m}\sin (m\tau
)p_{0}^{I}
\nonumber\\
&+& i\sum\limits_{n\neq 0}c_{n}e^{-i\omega _{n(0)}\tau }\cos \frac{%
k_{n}\sigma }{2}a_{n}^{I},  
\label{solution-open-nn} \\
x_{DD}^{I}(\tau ,\sigma ) &=&\frac{q_{1}^{I}\sinh (m\sigma )-q_{0}^{I}\sinh
(m(\sigma -\pi ))}{\sinh (m)}
\nonumber\\
&+&i\sum_{n\neq 0}\frac{1}{\sqrt{|\omega _{n(0)}|}}e^{-i\omega _{n(0)}\tau }\sin \frac{k_{n}\sigma }{2}a_{n}^{I},
\label{solution-open-dd} \\
x_{closed}^{I}(\tau ,\sigma ) &=&\cos (m\tau )x_{0}^{I}+\frac{1}{m}\sin
(m\tau )p_{0}^{I}
\nonumber\\
&+&i\sum\limits_{n\neq 0}c_{n}\left[ e^{-i(\omega _{n}\tau -k_{n}\sigma
)}a_{n}^{I}+e^{-i(\omega _{n}\tau +k_{n}\sigma )}\overline{a}_{n}^{I}\right]
.  
\label{solution-closed}
\end{eqnarray}%
Here, the following notations are used 
\begin{equation}
c_{n}=\frac{\mbox{sign}(n)}{\sqrt{|\omega _{n(0)}|}}~~,~~
\omega _{\pm |n|}=\pm \sqrt{k_{n}^{2}+m^{2}}~~,~~\omega _{\pm |n|(0)}=\pm 
\sqrt{{\frac{k_{n}}{2}}^{2}+m^{2}}~~,~~k_{n}=2\pi n,  \label{omegas}
\end{equation}%
The coordinates of the open string ends with D-D boundary
conditions are: $x_{DD}^{I}(\sigma =0)=q_{0}$ and $x_{DD}^{I}(\sigma =\pi
)=q_{1}$, respectively. The Fourier expansions of the general solutions of
the equations of motion for the fermionic fields are \cite{rrm,bp,dp} 
\begin{eqnarray}
S^{a}(\tau ,\sigma ) &=&\cos (m\tau )S_{0}^{a}+\sin (m\tau )\left( \Pi 
\overline{S}_{0}\right) ^{a} 
\nonumber\\
&+&\sum\limits_{n\neq 0}d_{n}\left[ e^{-i(\omega _{n}\tau -k_{n}\sigma
)}S_{n}^{a}+i\frac{\omega _{n}-k_{n}}{m}e^{-i(\omega _{n}\tau +k_{n}\sigma
)}\left( \Pi \overline{S}_{n}\right) ^{a}\right] , 
\label{solution-fermion1}\\
\overline{S}^{a}(\tau ,\sigma ) &=&\cos (m\tau )\overline{S}_{0}^{a}-\sin
(m\tau )\left( \Pi S_{0}\right) ^{a} 
\nonumber\\
&+&\sum\limits_{n\neq 0}d_{n}\left[ e^{-i(\omega _{n}\tau -k_{n}\sigma )}%
\overline{S}_{n}^{a}-i\frac{\omega _{n}-k_{n}}{m}e^{-i(\omega _{n}\tau
+k_{n}\sigma )}\left( \Pi S_{n}\right) ^{a}\right] ,
\label{solution-fermion2}
\end{eqnarray}
where $d_{n}=\left[ 1+\left( \frac{\omega _{n}-k_{n}}{m}\right) ^{2}\right]
^{-1/2}$.

In order to quantize the theory, the equal-time Poisson-Dirac (anti)brackets of the superstring fields and their conjugat momenta are required to hold \cite{rrm,mt}
\bea
\left[ x^{I}(\tau ,\sigma ),p^{J}(\tau ,\sigma ^{\prime })\right] &=& 
i\delta^{IJ}\delta (\sigma -\sigma ^{\prime })
,\label{bosonic-equal-time}\\
\left\{ S^{a}(\tau ,\sigma ),S^{b}(\tau ,\sigma ^{\prime })\right\} &=& 
\delta^{ab}\delta (\sigma -\sigma ^{\prime })
.\label{fermionic-equal-time}
\eea
The above (anti)brackets imply the following non-zero commutation relations for the superstring modes
\begin{eqnarray}
\left[ a_{m}^{I},a_{n}^{J}\right]  & = & \left[ \overline{a}_{m}^{I},\overline{a}_{n}^{J}\right] = 
\mbox{sign}(m)\delta _{m+n,0}\delta ^{IJ}, 
\label{comm-bosons}\\
\left\{ S_{m}^{a},S_{n}^{b}\right\}  & = & \left\{ \overline{S}_{m}^{a}, \overline{S}_{n}^{b}\right\} = 
\delta _{m+n,0}\delta ^{ab}.
\label{anticomm-fermions}
\end{eqnarray}
The zero-modes are conveniently combined in creation and annihilation
operators as follows 
\begin{eqnarray}
a_{0}^{I} & = & \frac{1}{\sqrt{2m}}\left( p_{0}^{I}-imx_{0}^{I}\right),a_{0}^{I\dag } = 
\frac{1}{\sqrt{2m}}\left( p_{0}^{I}+imx_{0}^{I}\right) ,
\label{zero-modes-bosons}\\
S_{\pm }^{a} & = & \frac{1}{2}\left( 1\pm \Pi \right) _{b}^{a}\frac{1}{\sqrt{2m}}
\left( S_{0}^{b}\pm i\overline{S}_{0}^{b}\right) ,S_{\pm }^{a\dag } = 
\frac{1}{2}\left( 1\pm \Pi \right) _{b}^{a}\frac{1}{\sqrt{2m}}\left( S_{0}^{b}\mp i
\overline{S}_{0}^{b}\right).
\label{zero-modes-fermions} 
\end{eqnarray}
With the redefinitions expressed by the relations (\ref{zero-modes-bosons}) and (\ref{zero-modes-fermions}), the (anti)commutators (\ref{comm-bosons}) and (\ref{anticomm-fermions}) can be seen to hold for $ n = 0, \pm 1, \pm 2\ldots$ for bosons and $ n = \pm , \pm 1, \pm 2, \ldots $ for fermions, respectively. The light-cone superstring hamiltonian density takes the form
\bea
P^{-}&=&-H_{l.c.}=\frac{1}{2p^{+}}\left(
p_{0}^{2}+m^{2}x_{0}^{2}+2iS_{0}^{a}\Pi _{ab}\overline{S}_{0}^{b}\right) +
\nonumber\\
&&\frac{1}{p^{+}}\sum_{n=1}^{\infty }\omega _{n}\left( a_{n}^{I\dag }a_{n}^{I}+%
\overline{a}_{n}^{I\dag }\overline{a}_{n}^{I}+S_{n}^{a\dag }S_{n}^{a}+%
\overline{S}_{n}^{a\dag }\overline{S}_{n}^{a}\right) .
\label{light-cone-Hamiltonian}
\eea
The superstring states in the Fock space representation can be constructed in the usual manner from the superstring vacuum state which is defined as follows
\begin{eqnarray}
a_{0}^{I}\left\vert 0\right\rangle  &=& a_{n}^{I}\left\vert 0\right\rangle =
\overline{a}_{n}^{I}\left\vert 0\right\rangle =0 ~~,~~n=1,2,3,\ldots 
\label{vacuum1} \\
\left( S_{0}^{a}+i\overline{S}_{0}^{a}\right) \left\vert 0\right\rangle 
& = & S_{n}^{a}\left\vert 0\right\rangle =\overline{S}_{n}^{a}\left\vert
0\right\rangle = 0 ~~,~~n=1,2,3,\ldots    \label{vacuum2} 
\end{eqnarray}
The vacuum state in the spinor sector has a similar structure to the vacuum of the type IIB GS superstring in the flat spacetime. In particular, its zero mode component describe massless spinors in the chiral representation of $SO(8)$ \cite{rrm,mt}. The generic Fock space vectors are built up by acting with the creation operators upon the vacuum
\begin{equation}
\left\vert \Psi \right\rangle = 
\Psi \left( a_{0}^{\dag },a_{-n},S_{0},S_{-n},\overline{a}_{0}^{\dagger},\overline{a}_{-n},
\overline{S}_{0},\overline{S}_{-n}\right) \left\vert 0\right\rangle ,  
\label{generic-vector-in-Fock}
\end{equation}
for $n >0$. However, the physical space is a subspace of the Fock space obtained by imposing the Virasoro constraints in their operatorial form, which are reduced to the level-matching conditions in the light-cone gauge
\be
\left( N-\overline{N}\right) \left\vert \Psi _{phys}\right\rangle  =0 ,
\label{level-matching-condition}
\ee
where
\be
N =\sum_{n=1}^{\infty }k_{n}\left( a_{-n}^{I}a_{n}^{I}+S_{-n}^{a}S_{n}^{a}\right) ~~,~~  
\overline{N} =\sum_{n=1}^{\infty }k_{n}\left( \overline{a}_{-n}^{I}
\overline{a}_{n}^{I}+\overline{S}_{-n}^{a}\overline{S}_{n}^{a}\right) .
\label{number-operators}
\ee
The quantum superstring is invariant under the kinematical supersymmetry generated by the set of charges
$\{ P^+, P^I, J^{+I}, J^{ij}, J^{i'j'}, Q^+, \overline{Q}^+ \}$ and the dynamical supersymmetry generated by
$\{ P^- , Q^- , \overline{Q}^- \}$, respectively \cite{rrm}. 

The boundary states in {\em pp}-wave background can be obtained by imposing the Dirichlet and Neumann boundary conditions in the bosonic sector of the open superstring and requiring that half of the background supersymmetry be preserved \cite{bp,dp,bgg}. Since it is possible to interpret the cylinder diagram in both open and closed string sectors, the open string boundary conditions can be mapped into the corresponding closed string boundary conditions by modular transformations as in the flat spacetime \cite{bgg}. Therefore, the boundary states can be constructed from the states of the Fock space of the closed superstring in a similar way to the Minkowski background case, with the important difference that there are coherent state contributions from zero modes \cite{bp,dp,bgg}. The boundary state conditions are given by the following relations
\begin{eqnarray}
\left( \partial _{+}x^{I}-M_{J}^{I}\partial _{-}x^{J}\right) \left\vert B\right\rangle  &=&0,  
\label{bosonic-bc} \\
\left( Q_{a}+i \eta M_{(s)ab}\overline{Q}_{b}\right) \left\vert B\right\rangle &=&0,  
\label{fermionic-bc-1} \\
\left( Q_{\dot{a}}+i \eta M_{(c)\dot{a}\dot{b}}\overline{Q}_{\dot{b}}\right) \left\vert B\right\rangle 
&=&0,
\label{fermionic-bc-2}
\end{eqnarray}
where $\eta =\pm 1$ corresponds to branes and anti-branes solutions, respectively, and $M_{IJ}$, $M_{(s)ab}$ and $M_{(c)\dot{a}\dot{b}}$ are constant matrices in the vectorial and chiral representations of $SO(8)$, respectively, whose form is dictated by the preserved supersymmetry. In particular, $M_{IJ} = \mbox{diag} \left( \pm 1, \pm 1, \ldots, \pm 1\right)$ with $-1$ for Neumann and $+1$ for Dirichlet boundary conditions, respectively.   By using the Fourier expansions of the superstring fields given in the relations (\ref{solution-closed}) - (\ref{solution-fermion2}), one can write the boundary conditions (\ref{bosonic-bc})-(\ref{fermionic-bc-2}) in terms of the creation and annihilation operators for superstring modes 
\begin{eqnarray}
\left( a_{n}^{I}-M^{I}_{J}\overline{a}_{n}^{J}\right) \left\vert B\right\rangle &=&0,  
\label{bosonic-operator-bc} \\
\left( S_{n}^{a}+i\eta M_{(s)b}^{a}\overline{S}_{n}^{b}\right) \left\vert B\right\rangle  &=&0,  \label{fermionic-operator-bc}
\end{eqnarray}
for all $n \in \mathbb{Z}$. The $D{p}$-brane boundary state that satisfy the above equations has the coherent state form \cite{bp,dp,bgg}
\begin{eqnarray}
\left\vert B\right\rangle  &=&\exp \sum_{n=1}^{\infty }\left(
M_{IJ}a_{n}^{I\dag }\overline{a}_{n}^{J\dag }-i\eta M_{(s)ab}S_{n}^{a}%
\overline{S}_{n}^{b\dag }\right) \left\vert B\right\rangle _{0},
\label{boundary-state-1} \\
\left\vert B\right\rangle _{0} &=&{\mathbf N}_{p+1}\left( M_{IJ}\left\vert
I\right\rangle \overline{\left\vert J\right\rangle }+i\eta M_{(c)\dot{a}\dot{%
b}}\left\vert \dot{a}\right\rangle \overline{\vert \dot{b}\rangle 
}\right) \exp \left( \frac{1}{2}M_{IJ}a_{0}^{I\dag }a_{0}^{J\dag }\right)
\left\vert 0\right\rangle _{a}.  
\label{boundary-state-2}
\end{eqnarray}
The normalization constant for a $D{p}$-brane was computed in \cite{bgg} from the static interaction amplitude between two-branes and has the form
\be
{\mathbf N}_{p+1} = \left( 2 \sinh \pi m \right)^{\frac{3-p}{2}}.
\label{normalization-constant}
\ee
The boundary states given in the relations (\ref{boundary-state-1})-(\ref{normalization-constant}) preserve sixteen supersymmetries for $ p = 3, 5, 7$ and for certain orientations of the world volume with respect to the light-cone directions. If the light-cone directions are transverse to the world-volume then the boundary states represent supersymmetric instantons for $p=1, 3, 5$. In general, the quantum states    
(\ref{boundary-state-1})-(\ref{normalization-constant}) correspond to curved branes and instantons \cite{bp,dp,bgg}.

\section{Thermalization of the GS superstring and thermal boundary states}

In this section we are going to construct the thermal boundary states from the type IIB GS superstring in the light-cone gauge in the {\em pp}-wave background. There is a similarity with the flat spacetime case which will be exploited to parallel the method developed in \cite{ivv1,ivv2,ivv3} and applied to the type IIB GS superstring in the Minkowski spacetime in \cite{ivv11}. This method is based on the TFD formalism in which the string at finite temperature is treated in the canonical quantization. Since the $D$-branes are build out of excitations from the Fock space, the TFD approach is particularly suited to discuss the boundary states.

\subsection{Total string at $T=0$}

In order to construct the boundary states at finite temperature, one has to obtain firstly the thermal string starting from the type IIB GS superstring. As in the flat spacetime, one can imagine that the superstring heats up as a consequence of its interaction with a thermal reservoir. In the TFD formalism, this process is the result of a particular interaction between each superstring mode and an identical copy of it which represents an oscillator degree of freedom of the reservoir. Thus, the relevant degrees of freedom of the reservoir pair the superstring modes. Therefore, they can be taken to form a copy of the original GS superstring which will be denoted by a tilde; the two copies form the total string. The interaction between the superstring and the tilde superstring takes the total string from $T = 0$ to $T \neq 0$ and represents the thermalization process.  Formally, the thermalization is implemented by a temperature dependent  unitary operator obtained by exponentiating the Bogoliubov operator of the total string. By this mapping, the thermal string at $ T \neq 0$ formally preserves the structure of the total string at $T = 0$, i. e. twice the structure of the superstring. However, the physical information is extracted by taking e. g. the expectation values of the superstring operators without tilde (not belonging to the reservoir) at $T = 0$ in the thermal states. That shows that the structure of the thermal string is different from the one of the original GS superstring. For example, the supersymmetry is broken \cite{ng1,ng2,ivv11}. Also, a thermal quantum excitation can be viewed as a mixture of superstring and tilde superstring quantum excitations at $T = 0$. 

In general, labelling one of the superstrings with tilde, i. e. choosing which copy represents the degrees of freedom of the thermal reservoir, is a symmetry operation of the total system. In the lagrangian formulation of the TFD, this symmetry can be implemented in the theory if the following lagrangian density is taken for the total system
\be
\hat{\LL}( \Phi, \tilde{\Phi} ) = \LL(\Phi ) - \tilde{\LL}(\tilde{\Phi}),
\label{lagrangian-density}
\ee 
where $\Phi$ and $\tilde{\Phi}$ denote a generic field of the total string. The $\tilde{\LL}(\tilde{\Phi})$ can be defined according to two different recipes: $\tilde{\LL}(\tilde{\Phi}) = [\LL (\Phi)]{\tilde{~}} $  and 
$\tilde{\LL}(\tilde{\Phi})= \LL (\tilde{\Phi})$, respectively. It was shown in \cite{ivv11} that for type IIB GS superstring in flat spacetime the two definitions are equivalent, i. e. taking the tilde-conjugation of the lagrangian ammounts to replacing $\Phi$ by $\tilde{\Phi}$. However, the situation is different in the present case. Indeed, by 
applying the relations from the Appendix A, the tilde-conjugation operation performed in {\em pp}-wave background  changes the sign of the last two terms of the lagrangian defined in the relation (\ref{action}). The reason for that is the representation of the Dirac matrices in $d=2$ which in the flat spacetime was taken to be $\rho^0 = \sigma_2, \rho^1=i\sigma_1$ while in the {\em pp}-wave background it is $\rho^0 = -i\sigma_2 , \rho^1 = \sigma_1$. If the representation were changed to the one used in the flat spacetime, the last fermionic term in the action (\ref{action}) would still change its sign upon tilde conjugation due to the fact that the fermions $S$ and $\overline{S}$ and the matrix $\Pi$ are real. The change of the relative sign between the bosonic and the fermionic lagrangians leads to a modification of the coefficients of the dynamical supersymmetry transformations (\ref{dyn-susy1}) - (\ref{dyn-susy3}) for the tilde string to $\a = \b = \o = \m = 1$ and $\l = \s = 1$. Another consequence of the sign change in the tilde lagrangian is the addition of the fermionic lagrangians in the relation (\ref{lagrangian-density}) and of the corresponding fermionic energies, too. Since that would uncharacterize the thermalization interaction, in what follows we will consider the definition of the total string action as given by (\ref{lagrangian-density}) and 
\be
\tilde{\LL}(\tilde{\Phi})= \LL (\tilde{\Phi}).
\label{tilde-lagrangian}
\ee
The representations of the Dirac matrices in two and ten dimensions are same as in the previous section \cite{rrm,mt}. Also, the dynamical supersymmetry transformations have the same coefficients for non-tilde as well as for tilde superstring. With this definition of the total lagrangian, the method of \cite{ivv11} applies straightforwardly to the type IIB GS superstring in the {\em pp}-wave background.

The equations of motion derived from (\ref{lagrangian-density}) by varying the superstring and the tilde superstring actions independently, represent two copies of the equations of motion given in the relations (\ref{eq-mot-bosons1})-(\ref{eq-mot-spinors3}) 
\begin{eqnarray}
\partial _{+}\partial _{-}x^{I}+m^{2}x^{I} &=&0~~,~~\partial _{+}\partial _{-}
\widetilde{x}^{I}+m^{2}\widetilde{x}^{I}=0
\label{eq-mot-total1} \\
\partial _{+}S^{a}-m(\Pi \overline{S})^{a} &=&0~~,~~\partial _{+}\widetilde{S}%
^{a}-m(\Pi \widetilde{\overline{S}})^{a}=0,
\label{eq-mot-total2} \\
\partial _{-}\overline{S}^{a}+m(\Pi S)^{a} &=&0~~,~~\partial _{-}\widetilde{%
\overline{S}}^{a}+m(\Pi \widetilde{S})^{a}=0 .
\label{eq-mot-total3}
\end{eqnarray}
Here, since the two copies are identical, we have taken $\tilde{m} = m$ and $\tilde{\Pi}=\Pi$. This choice is consistent with the tilde conjugation rule for real numbers. Similarly, the boundary conditions obtained from  (\ref{lagrangian-density}) represent two sets of independent boundary conditions for the superstring and the tilde superstring, respectively. However, since the two copies of the superstring are identical, the chirality of the superstring should be chosen to be the same as the chirality of the tilde superstring. Note that the total string is invariant under two sets of independent supersymmetry transformations.

The total string modes can be quantized by imposing the canonical quantization (anti)commutators in both superstring and tilde-superstring sectors which are just two copies of the (anti)commutation relations (\ref{comm-bosons}) and (\ref{anticomm-fermions}) 
\begin{eqnarray}
\left[ a_{m}^{I},a_{n}^{J}\right]  &=&\left[ \overline{a}_{m}^{I},\overline{a}_{n}^{J}\right] = 
\mbox{sign}(m)\delta _{m+n,0}\delta ^{IJ},
\label{comm-total-string-1} \\
\left[ \widetilde{a}_{m}^{I},\widetilde{a}_{n}^{J}\right]  &=& [\widetilde{\overline{a}}_{m}^{I},\widetilde{\overline{a}}_{n}^{J} ] =
\mbox{sign}(m)\delta _{m+n,0}\delta ^{IJ},  
\label{comm-total-string-2} \\
\left\{ S_{m}^{a},S_{n}^{b}\right\}  &=&\{ \overline{S}_{m}^{a},
\overline{S}_{n}^{b} \} =\delta _{m+n,0}\delta ^{ab},
\label{anticomm-total-string-1} \\
\{ \widetilde{S}_{m}^{a},\widetilde{S}_{n}^{b} \}  &=& 
\{ \widetilde{\overline{S}}_{m}^{a},\widetilde{\overline{S}}_{n}^{b} \} = \delta _{m+n,0}\delta ^{ab}.  
\label{anticomm-total-string-2} 
\end{eqnarray}
The Hilbert space of the total string is given by the tensor product $\hat{{\cal H}}={\cal H}\otimes \tilde{{\cal H}}$.  Let us denote by $|\Psi \rangle \rangle $ a generic vector from the total Hilbert space $\hat{{\cal H }}$. The total string excitations can be obtained by acting with the creation and annihilation operators from the non-tilde and tilde superstring sectors on the total vacuum state 
$|0\rangle \rangle =|0\rangle \widetilde{ |0\rangle }\in \hat{{\cal H}}$.\ The total vacuum state satisfies the
properties of the superstring vacuum (\ref{vacuum1}) and (\ref{vacuum2}) in each sector 
\begin{eqnarray}
a_{0}^{I}|0\rangle \rangle  &=&a_{n}^{I}|0\rangle \rangle =\overline{a}_{n}^{I}|0\rangle \rangle =0~~,~~n=1,2,3,\ldots ,
\label{total-vacuum-state-1} \\
\widetilde{a}_{0}^{I}|0\rangle \rangle  &=&\widetilde{a}_{n}^{I}|0\rangle \rangle = \widetilde{\overline{a}}_{n}^{I}|0\rangle \rangle = 0~~,~~n=1,2,3,\ldots ,  
\label{total-vacuum-state-2} \\
\left( S_{0}^{a}+i\overline{S}_{0}^{a}\right) |0\rangle \rangle  &=& S_{n}^{a}|0\rangle \rangle = \overline{S}_{n}^{a}|0\rangle \rangle = 0~~,~~n=1,2,3,\ldots ,  
\label{total-vacuum-state-3} \\
( \widetilde{S}_{0}^{a}+i\widetilde{\overline{S}}_{0}^{a} )
|0\rangle \rangle  &=&\widetilde{S}_{n}^{a}|0\rangle \rangle = 
\widetilde{\overline{S}}_{n}^{a}|0\rangle \rangle =0~~,~~n=1,2,3,\ldots ,
\label{total-vacuum-state-4}
\end{eqnarray}
where each operator acts on its sector from $|0\rangle \rangle $, e. g. 
$a_{n}^{I}|0\rangle \rangle =(a_{n}^{I}|0\rangle )\widetilde{|0\rangle }=0$.
The total ground state $|\phi _{0}\rangle \rangle \in \hat{{\cal H}}$ can be obtained as the tensor product of the supersymmetric ground state of non tilde and tilde superstrings. As in the flat spacetime \cite{ivv11}, the ground state of the total string can be written in the following form 
\begin{eqnarray}
|\phi _{0}\rangle \rangle  & = & 
\mbox{diag}(|\phi _{0}\rangle |\tilde{\phi}_{0}\rangle ),  
\label{total-ground-state-1a} \\
\left( |\phi _{0}\rangle \rangle \right) \tilde{} & = & 
|\phi _{0}\rangle \rangle .  
\label{total-ground-state-2a}
\end{eqnarray}
The ground state can be decomposed into states that are direct products of vacua from various string sectors 
\begin{equation}
|\phi _{0}\rangle \rangle = 
\{ |I\rangle |\overline{J}\rangle |\tilde{I} \rangle |\tilde{\overline{J}}\rangle ,|a\rangle |\overline{J}\rangle |\tilde{a}\rangle |\tilde{\overline{J}}\rangle ,|I\rangle |\overline{b}\rangle |
\tilde{I}\rangle |\tilde{\overline{b}}\rangle ,|a\rangle |\overline{b}
\rangle |\tilde{a}\rangle |\tilde{\overline{b}}\rangle \}\equiv \{|\phi
\rangle \rangle |\bar{\phi}\rangle \rangle \},  
\label{total-ground-state}
\end{equation}
where $|\phi \rangle \rangle =|\phi \rangle |\tilde{\phi}\rangle $ and 
$| \bar{\phi}\rangle \rangle =|\bar{\phi}\rangle |\tilde{\bar{\phi}}\rangle $,
and $\phi =\{I,a\}$. The state (\ref{total-ground-state}) is tilde-invariant and consistent with the Kronecker product \cite{ivv11}. A general vector from the Fock space of the total string is obtained by acting with the creation operators on to the vacuum state
\begin{equation}
\left\vert \Psi \right\rangle \rangle =\Psi \left( a_{0}^{\dag },a_{-n},\overline{a}_{0}^{\dagger },
\overline{a}_{-n}, \widetilde{a}_{0}^{\dag }, \widetilde{a}_{-n},
\widetilde{\overline{a}}_{0}^{\dagger },\widetilde{\overline{a}}_{-n},
S_{0},S_{-n}, \overline{S}_{0},\overline{S}_{-n}, 
\widetilde{S}_{0},\widetilde{S}_{-n},\widetilde{\overline{S}}_{0}, \widetilde{\overline{S}}_{-n}\right) 
\left\vert 0\right\rangle \rangle .
\label{general-total-state}
\end{equation}
The physical states are the states (\ref{general-total-state}) that satisfy
the level matching condition in each sector 
\bea
\left( N-\overline{N}\right) \left\vert \Psi _{phys}\right\rangle \rangle =
\left( \widetilde{N}-\widetilde{\overline{N}}\right) \left\vert \Psi_{phys}\right\rangle \rangle =0 ,
\eea
where 
\begin{eqnarray}
N &=&\sum_{n=1}^{\infty } k_{n}\left(a_{-n}^{I}a_{n}^{I}+S_{-n}^{a}S_{n}^{a}\right) ~~,~~
\overline{N} =\sum_{n=1}^{\infty } k_{n}\left( \overline{a}_{-n}^{I}\overline{a}_{n}^{I}+
\overline{S}_{-n}^{a}\overline{S}_{n}^{a}\right) , 
\label{total-number-operator-1}\\
\widetilde{N} &=&\sum_{n=1}^{\infty } k_{n}
\left( \widetilde{a}_{-n}^{I} \widetilde{a}_{n}^{I}+\widetilde{S}_{-n}^{a}\widetilde{S}_{n}^{a}\right)
~~,~~\widetilde{\overline{N}}=\sum_{n=1}^{\infty } k_{n} 
\left( \widetilde{ \overline{a}}_{-n}^{I}\widetilde{\overline{a}}_{n}^{I}+
\widetilde{\overline{S}}_{-n}^{a}\widetilde{\overline{S}}_{n}^{a}\right) .
\label{total-number-operator-2}
\end{eqnarray}
Here, we have taken $k_{n}=\overline{k_{n}}=\widetilde{k_{n}}=\widetilde{\overline{k_{n}}}$ because these factors should be the same for identical modes. That results from the tilde conjugation rules for real numbers, too. 

By varying the lagrangian density (\ref{lagrangian-density}) with respect to the non tilde and tilde fields one obtains two independent sets of boundary conditions that preserve half of the supersymmetry in each sector and define the total $D$-brane boundary states from $\hat{\HH}$. They are given by a copy of the relations (\ref{bosonic-bc})-(\ref{fermionic-bc-2}) in each sector
\begin{eqnarray}
\left( \partial _{+}x^{I}-M_{J}^{I}\partial _{-}x^{J}\right) \left\vert
B\right\rangle \rangle  &=&0,\left( \partial _{+}\widetilde{x}^{I}-%
\widetilde{M}_{J}^{I}\partial _{-}\widetilde{x}^{J}\right) \left\vert
B\right\rangle \rangle =0,  \label{bosonic-bc-total-string} \\
\left( Q_{a}+i\eta M_{(s)ab}\overline{Q}_{b}\right) \left\vert
B\right\rangle \rangle  &=&0,\left( \widetilde{Q}_{a}+i\widetilde{\eta }%
\widetilde{M}_{(s)ab}\widetilde{\overline{Q}}_{b}\right) \left\vert
B\right\rangle \rangle =0,  \label{fermionic-bc-total-string-1} \\
\left( Q_{\dot{a}}+i\eta M_{(c)\dot{a}\dot{b}}\overline{Q}_{\dot{b}}\right)
\left\vert B\right\rangle \rangle  &=&0,\left( \widetilde{Q}_{\dot{a}}+i%
\widetilde{\eta }\widetilde{M}_{(c)\dot{a}\dot{b}}\widetilde{\overline{Q}}_{%
\dot{b}}\right) \left\vert B\right\rangle \rangle =0.
\label{fermionic-bc-total-string-2}
\end{eqnarray}
Consequently, the Fock space boundary conditions of the total string are
\begin{eqnarray}
\left( a_{n}^{I}-M_{J}^{I}\overline{a}_{n}^{J}\right) \left\vert
B\right\rangle \rangle  &=&0,\left( \widetilde{a}_{n}^{I}-\widetilde{M}%
_{J}^{I}\widetilde{\overline{a}}_{n}^{J}\right) \left\vert B\right\rangle
\rangle =0,  
\label{total-boundary-conditions-Fock-1} \\
\left( S_{n}^{a}+i\eta M_{(s)b}^{a}\overline{S}_{n}^{b}\right) \left\vert
B\right\rangle \rangle  &=&0,\left( \widetilde{S}_{n}^{a}+i\widetilde{\eta }%
\widetilde{M}_{(s)b}^{a}\widetilde{\overline{S}}_{n}^{b}\right) \left\vert
B\right\rangle \rangle =0.  
\label{total-boundary-conditions-Fock-2}
\end{eqnarray}%
Note that in the relations (\ref{bosonic-bc-total-string})-(\ref%
{total-boundary-conditions-Fock-2}) the matrices $M$ and $\widetilde{M}$ do
not need to be identical.\ Actually, they coincide up to a similarity
transformation that satisfies the conditions obtained from the requirement
of preserving sixteen supercharges \cite{bp}. Also, $\eta $ and $\widetilde{%
\eta }$ can, in principle, be chosen differently in each sector since the
TFD\ does not give any prescription regarding the boundary states. However,
if one sticks to the idea that the two copies of the system be completly
identical, then one should take $M=\widetilde{M}$ and $\eta =\widetilde{\eta 
}$. Due to the form of the boundary conditions  (\ref%
{bosonic-bc-total-string})-(\ref{total-boundary-conditions-Fock-2}), the
total boundary state $\left\vert B\right\rangle \rangle $ can be factorized
as
\be
\left\vert B\right\rangle \rangle =\left\vert B\right\rangle \widetilde{%
\left\vert B\right\rangle } ,
\label{factorized-boundary-state}
\ee 
where
\begin{eqnarray}
\left\vert B\right\rangle  &=&\exp \sum_{n=1}^{\infty }\left(
M_{IJ}a_{n}^{I\dag }\overline{a}_{n}^{J\dag }-i\eta M_{(s)ab}S_{n}^{a}%
\overline{S}_{n}^{b\dag }\right) \left\vert B\right\rangle _{0},
\label{total-boundary-states-1} \\
\widetilde{\left\vert B\right\rangle } &=&\exp \sum_{n=1}^{\infty }\left( 
\widetilde{M}_{IJ}\widetilde{a}_{n}^{I\dag }\widetilde{\overline{a}}%
_{n}^{J\dag }-i\widetilde{\eta }\widetilde{M}_{(s)ab}\widetilde{S}_{n}^{a}%
\widetilde{\overline{S}}_{n}^{b\dag }\right) \widetilde{\left\vert
B\right\rangle }_{0},  \label{total-boundary-states-2} \\
\left\vert B\right\rangle _{0} &=&{\bf N}_{p+1}\left( M_{IJ}\left\vert
I\right\rangle \overline{\left\vert J\right\rangle }+i\eta M_{(c)\dot{a}\dot{%
b}}\left\vert \dot{a}\right\rangle \overline{|\dot{b}\rangle }\right) \exp
\left( \frac{1}{2}M_{IJ}a_{0}^{I\dag }a_{0}^{J\dag }\right) \left\vert
0\right\rangle _{a},  \label{total-boundary-states-3} \\
\widetilde{\left\vert B\right\rangle }_{0} &=&\widetilde{{\bf N}}%
_{p+1}\left( \widetilde{M}_{IJ}\widetilde{\left\vert I\right\rangle }%
\widetilde{\overline{\left\vert J\right\rangle }}+i\widetilde{\eta }%
\widetilde{M}_{(c)\dot{a}\dot{b}}\widetilde{\left\vert \dot{a}\right\rangle }%
\widetilde{\overline{|\dot{b}\rangle }}\right) \exp \left( \frac{1}{2}%
\widetilde{M}_{IJ}\widetilde{a}_{0}^{I\dag }\widetilde{a}_{0}^{J\dag
}\right) \widetilde{\left\vert 0\right\rangle }_{a}.
\label{total-boundary-states-4}
\end{eqnarray}%
A remark is in order here. In the relations  (\ref{bosonic-bc-total-string})-(\ref{total-boundary-states-4}) the tilde over any object does not necessarily correspond to the tilde conjugation, although in some cases may coincide to
it as for ${\bf N}_{p+1} = \widetilde{{\bf N}}_{p+1}$. Tilde is merely a label to denote reservoir 
quantities that are defined by identical equations due to the definition of the total lagrangian given in 
(\ref{lagrangian-density}) and (\ref{tilde-lagrangian}). These two relations also ensure that the total hamiltonian is $\widehat{H}_{l.c.}= H_{l.c.}-\widetilde{H}_{l.c.}$, 
where 
\begin{eqnarray}
-H_{l.c.} &=&\frac{1}{2p^{+}}\left( p_{0}^{2}+m^{2}x_{0}^{2}+2iS_{0}^{a}\Pi
_{ab}\overline{S}_{0}^{b}\right) 
\nonumber\\
&+&\frac{1}{p^{+}}\sum_{n=1}^{\infty }\omega
_{n}\left( a_{n}^{I\dag }a_{n}^{I}+\overline{a}_{n}^{I\dag }\overline{a}%
_{n}^{I}+S_{n}^{a\dag }S_{n}^{a}+\overline{S}_{n}^{a\dag }\overline{S}%
_{n}^{a}\right) ,  
\label{total-hamiltonian-1} \\
-\widetilde{H}_{l.c.} &=&\frac{1}{2\widetilde{p}^{+}}\left( \widetilde{p}%
_{0}^{2}+m^{2}\widetilde{x}_{0}^{2}+2i\widetilde{S}_{0}^{a}\widetilde{\Pi }%
_{ab}\widetilde{\overline{S}}_{0}^{b}\right) \nonumber\\ 
&+& \frac{1}{\widetilde{p}^{+}}%
\sum_{n=1}^{\infty }\omega _{n}\left( \widetilde{a}_{n}^{I\dag }\widetilde{a}%
_{n}^{I}+\widetilde{\overline{a}}_{n}^{I\dag }\widetilde{\overline{a}}%
_{n}^{I}+\widetilde{S}_{n}^{a\dag }\widetilde{S}_{n}^{a}+\widetilde{%
\overline{S}}_{n}^{a\dag }\widetilde{\overline{S}}_{n}^{a}\right) .
\label{total-hamiltonian-2}
\end{eqnarray}%
Thus, the doubling of the original superstring to obtain the total string
corresponds effectively to taking two identical copies of the superstring
equations and boundary conditions.

\subsection{Thermal string at $T \neq 0$}

The total string modes from the non tilde and tilde sector interact among thermselves in a particular manner. As a result, the superstring heats to $ T \neq 0$. That is the TFD picture of superstring heating \cite{ubook}. It corresponds to the physical process in which the superstring is in contact with the thermal reservoir at thermodynamic equilibrium. The specific interaction responsible for the superstring heating can be described in terms of the temperature dependent Bogoliubov operator $G$ constructed from all non tilde and tilde modes. It can also be viewed as a map from the Hilbert space of the total string to a new Hilbert space corresponding to the thermal string. If one denotes by 
\bea
\OO &=& \{ O \} \equiv  \{ a_0, a^{\dagger}_{0}, \tilde{a}_0, \tilde{a}^{\dagger}_{0};\ 
a^{I}_n, a^{I \dagger}_n, \tilde{a}^{I}_n, \tilde{a}^{I\dagger}_n; \ 
\bar{a}^{I}_n, \bar{a}^{I \dagger}_n, \tilde{\bar{a}}^{I}_n, \tilde{\bar{a}}^{I\dagger}_n;\ 
\nonumber\\
&~& 
S_{\pm}, S^{\dagger}_{\pm}, \tilde{S}_{\pm}, \tilde{S}^{\dagger}_{\pm};\ 
S^{a}_n, S^{a \dagger}_n, \tilde{S}^{a}_n, \tilde{S}^{a\dagger}_n;\ 
\bar{S}^{a}_n, \bar{S}^{a \dagger}_n, \tilde{\bar{S}}^{a}_n, \tilde{\bar{S}}^{a\dagger}_n \},
\label{all-oscillators}
\eea
the set of all creation and annihilation operators, then the map generated by $G$ acts on $\OO$ by similarity transformations \cite{ivv12}
\be
\OO (\beta_T) = e^{-iG}\, \OO \, e^{iG} \equiv \{ e^{-iG}\, O \, e^{iG} \}.
\label{map-oscillators}
\ee
The above mapping represents the thermalization of the superstring. As in the flat spacetime, the temperature dependent operators $\OO (\beta_T)$ can be interpreted as corresponding to the Fourier modes of some thermal fields that satisfy the superstring equations of motion \cite{ivv12}
\bea
X^{I}(\b_T ) = e^{-iG}\, X^{I} \, e^{iG}~ , ~ S^{a}(\b_T ) = e^{-iG}\, S^{a} \, e^{iG}~,~ \bar{S}^{a}(\b_T ) = e^{-iG}\, \bar{S}^{a} \, e^{iG}, 
\label{thermal-string-operators}\\
\tilde{X}^{I}(\b_T ) = e^{-iG}\, \tilde{X}^{I} \, e^{iG} ~,~ \tilde{S}^{a}(\b_T ) = e^{-iG}\, \tilde{S}^{a} \, e^{iG} ~,~ \tilde{\bar{S}}^{a}(\b_T ) = e^{-iG}\, \tilde{\bar{S}}^{a} \, e^{iG}, 
\label{thermal-tilde-string-operators}
\eea
which can be obtained from a thermalized lagrangian density
\be
\LL(\b_T ) = e^{-iG}\hat{\LL}e^{iG}.
\label{thermal-lagrangian}
\ee
The thermal string excitations are obtained by acting with the thermal creation operators from the set $\OO (\beta_T)$ on the thermal vacuum state
\be
|0(\b_T )\rangle\rangle = e^{-iG}|0\rangle\rangle~,~|0\rangle\rangle = |0\rangle|\tilde{0}\rangle.
\label{thermal-vacuum}
\ee
The thermal ground state given in the relation (\ref{total-ground-state}) can be mapped to $T \neq 0$ to obtain the thermal ground state
\be
|\phi_{0}(\b_T)\rangle\rangle = e^{-iG}|\phi_0\rangle\rangle \equiv e^{-iG}\{ |\phi\rangle\rangle
|\bar{\phi}\rangle\rangle \} 
\label{thermal-ground-state}.
\ee 
These relations show that the thermalization process is the same as in the flat spacetime since its derivation relies upon the general properties of the Bogoliubov operator which, for the type IIB GS superstring in {\em pp}-wave background, has the following form \cite{ng1}
\begin{equation}
G = G^{B} + G^{F}, 
\label{bogoliubov-operator}
\end{equation}
where the bosonic $G^{B}$ and the fermionic $G^{F}$ operators, respectively, have the following general form
\bea
G^{B} & =& G_{0}^{B} + \sum_{n=1} \left( G_{n}^{B} + {\bar G}_n^{B} \right),
\label{bogoliubov-bosons}
\\
G^{F}&=&G_{+}^{F} + G_{-}^{F} + \sum_{n=1} \left( G_{n}^{F} + {\bar G}_{n}^{F}\right).
\label{bogoliubov-fermions}
\eea
The operators $G^{B}$ and $G^{F}$ represent the thermal interaction among the pairs of identical modes of the superstring and the thermal reservoir and their explicit form in terms of creation and annihilation operators is \cite{ubook}
\begin{eqnarray}
G_{0}^{B} &=&-i\theta _{0}^{B}(\beta _{T})\sum_{I=1}^{8}\left( a_{0}^{I}{%
\tilde{a}}_{0}^{I}-{\tilde{a}}_{0}^{I\dagger }a_{0}^{I\dagger }\right) ,
\label{bogoliubov-boson-zero} \\
G_{n}^{B} &=&-i\theta _{n}^{B}(\beta _{T})\sum_{I=1}^{8}\left( a_{n}^{I}{%
\tilde{a}}_{n}^{I}-{\tilde{a}}_{n}^{I\dagger } a_{n}^{I\dagger }\right)
,  \label{bogoliubov-boson-left} \\
{\bar{G}}_{n}^{B} &=&-i{\bar{\theta}}_{n}^{B}(\beta
_{T})\sum_{I=1}^{8}\left( \overline{{a}}_{n}^{I}\widetilde{\overline{a}}%
_{n}^{I}-\widetilde{\overline{a}}_{n}^{I\dagger }\overline{{a}}_{n}^{I\dag
}\right) ,  \label{bogoliubov-boson-right} \\
G_{\pm }^{F} &=&-i\theta _{\pm }^{F}(\beta _{T})\sum_{a=1}^{8}\left( 
\widetilde{S}_{\pm }^{a}S_{\pm }^{a}-S_{\pm }^{a\dagger }\widetilde{S}_{\pm
}^{a\dagger }\right) ,  \label{bogoliubov-fermion-zero} \\
G_{n}^{F} &=&-i\theta _{n}^{F}(\beta _{T})\sum_{a=1}^{8}\left( \widetilde{S}%
_{n}^{a}S_{n}^{a}-S_{n}^{a\dagger }\widetilde{S}_{n}^{a\dagger }\right) ,
\label{bogoliubov-fermion-left} \\
G_{n}^{F} &=&-i{\bar{\theta}}_{n}^{F}(\beta _{T})\sum_{a=1}^{8}\left( 
\widetilde{\overline{S}}_{n}^{a}\overline{S}_{n}^{a}-\overline{S}%
_{n}^{a\dagger }\widetilde{\overline{S}}_{n}^{a\dagger }\right) .
\label{bogoliubov-fermion-right}
\end{eqnarray}
The parameters $\theta^B$ and $\theta^F$ depend on the temperature and the statistics of the corresponding modes as follows \cite{ng1} 
\begin{eqnarray}
\theta _{0}^{B} &=&{\mbox{arcsinh}}\left( e^{\frac{\beta m}{p+}}-1\right) ^{-%
\frac{1}{2}},\qquad {\theta }_{\pm }^{F}=\arcsin \left( e^{\frac{\beta m}{p+}%
}-1\right) ^{-\frac{1}{2}},  \label{theta-zero-modes} \\
\theta _{n}^{B} &=&{\mbox{arcsinh}}\left( e^{\frac{\beta \omega _{n}}{p+}%
+i\lambda k_{n}}-1\right) ^{-\frac{1}{2}},\qquad \overline{\theta }_{n}^{B}={%
\mbox{arcsinh}}\left( e^{\frac{\beta \omega _{n}}{p+}-i\lambda
k_{n}}-1\right) ^{-\frac{1}{2}}, 
\label{theta-bosons}\\
\theta _{n}^{F} &=&\arcsin \left( e^{\frac{\beta \omega _{n}}{p+}+i\lambda
k_{n}}+1\right) ^{-\frac{1}{2}},\qquad \overline{\theta }_{n}^{F}=\arcsin
\left( e^{\frac{\beta \omega _{n}}{p+}-i\lambda k_{n}}+1\right) ^{-\frac{1}{2%
}}.  
\label{theta-fermions}
\end{eqnarray}
From these relations one can easily check that the thermal string operators formally satisfy the same supersymmetry algebra in both tilde and non tilde sectors as the total string. However, according to the TFD formalism, the symmetries should be checked for the non tilde operators at $T = 0$ against the thermal string states at $T \neq 0$. It is a simply exercise to show that the supersymmetries are broken.

\subsection{Thermal boundary states at $T \neq 0$}

Let us denote the boundary operators for the total string given in the
relations (\ref{total-boundary-conditions-Fock-1})-(\ref%
{total-boundary-conditions-Fock-2}) by 
\begin{eqnarray}
{\DD}_{n}^{I} &=&D_{n}^{I}=\left\{ a_{n}^{I}-M_{J}^{I}\overline{a}_{n}^{J},%
\widetilde{a}_{n}^{I}-\widetilde{M}_{J}^{I}\widetilde{\overline{a}}%
_{n}^{J}\right\} ,  \label{total-bc-to-therm-bos} \\
{\DD}_{n}^{a} &=&D_{n}^{a}=\left\{ S_{n}^{a}+i\eta M_{(s)b}^{a}\overline{S}%
_{n}^{b},S_{n}^{a}+i\eta M_{(s)b}^{a}\overline{S}_{n}^{b}\right\} .
\label{total-bc-to-therm-ferm}
\end{eqnarray}%
Then from the thermalization given by the relation \ref{map-oscillators} it follows that the thermal boundary operators have the following general form
\begin{eqnarray}
{\DD}_{n}^{I}(\beta _{T}) &=&e^{-iG}\,{\DD}_{n}^{I}\,e^{iG}\equiv
\{e^{-iG}{\em D}_{n}^{I}\,e^{iG}\},  
\label{thermal-bc-bosons} \\
{\DD}_{n}^{a}(\beta _{T}) &=&e^{-iG}\,{\DD}_{n}^{a}\,e^{iG}\equiv
\{e^{-iG}{\em D}_{n}^{a}\,e^{iG}\}.  
\label{thermal-bc-fermions}
\end{eqnarray}
These boundary conditions can be obtained by varying the lagrangian (\ref{thermal-lagrangian}) with respect to the worldsheet variables $\s^A$, too. This can be easily seen by noting that the thermalization does not affect the wave function, but only the creation and annihilation operators. It follows from (\ref{thermal-bc-bosons}) and (\ref{thermal-bc-fermions}) that the thermal boundary conditions in the thermal Hilbert space $\HH(\b_T)$
\begin{equation}
{\cal D}_{n}^{I}(\beta _{T})\left\vert B(\beta _{T})\right\rangle \rangle =%
{\cal D}_{n}^{a}(\beta _{T})\left\vert B(\beta _{T})\right\rangle \rangle =0,
\label{thermal-bc-problem}
\end{equation}
are satisfied by a thermal boundary state of the form
\be
\left\vert B(\beta _{T})\right\rangle \rangle = e^{-iG}\left\vert B \right\rangle \rangle ,
\label{thermal-boundary-state}
\ee
where $\left\vert B \right\rangle \rangle$ was given in the relations (\ref{factorized-boundary-state})-(\ref{total-boundary-states-4}). After a simple algebra, one can show that the thermal boundary state can be written in terms of the thermal vacuum state as
\be
\left\vert B(\beta _{T})\right\rangle \rangle ={\bf N}_{p+1}^{2}\exp \left(
\Sigma (\beta _{T})+\widetilde{\Sigma }(\beta _{T})\right)
\left\vert B(\beta _{T})\right\rangle \rangle _{0},
\label{thermal-boundary-state-final}
\ee
where 
\begin{eqnarray}
\Sigma (\beta _{T}) &=&\sum_{n=1}^{\infty }\left( M_{IJ}a_{n}^{I\dag }(\beta
_{T})\overline{a}_{n}^{J\dag }(\beta _{T})-i\eta M_{(s)ab}S_{n}^{a}(\beta
_{T})\overline{S}_{n}^{b\dag }(\beta _{T})\right) ,  
\label{sigma-1}\\
\widetilde{\Sigma }(\beta _{T}) &=&\sum_{n=1}^{\infty }\left( \widetilde{M}%
_{IJ}\widetilde{a}_{n}^{I\dag }(\beta _{T})\widetilde{\overline{a}}%
_{n}^{J\dag }(\beta _{T})-i\widetilde{\eta }\widetilde{M}_{(s)ab}\widetilde{S%
}_{n}^{a}(\beta _{T})\widetilde{\overline{S}}_{n}^{b\dag }(\beta
_{T})\right) .
\label{sigma-2}
\end{eqnarray}
The thermal zero mode state $\left\vert B(\beta _{T})\right\rangle \rangle
_{0}$ can be written in terms of the thermal ground state as follows 
\begin{eqnarray}
\left\vert B(\beta _{T})\right\rangle \rangle _{0} &=&
\left( M_{IJ} \vert I\overline{J}(\beta _{T}) \rangle \rangle + 
i\eta M_{(c)\dot{a}\dot{b}} \vert \dot{a}\overline{\dot{b}}(\beta _{T}) \rangle \rangle \right) 
\left( \widetilde{M}_{IJ}\vert \widetilde{I}\widetilde{\overline{J}}(\beta _{T})\rangle \rangle + 
i\widetilde{\eta }\widetilde{M}_{(c)\dot{a}\dot{b}}\vert \widetilde{\dot{a}}\widetilde{\overline{\dot{b}}}(\beta _{T})\rangle \rangle \right)  
\nonumber\\
&&\exp \left( \frac{1}{2}M_{IJ}a_{0}^{I\dag }(\beta_T)a_{0}^{J\dag }(\beta_T)\right) 
\exp
\left( \frac{1}{2}\widetilde{M}_{IJ}\widetilde{a}_{0}^{I\dag }(\beta_T)(\beta_T)\widetilde{a}_{0}^{J\dag }(\beta_T) 
\right) 
\left\vert 0(\beta _{T})\right\rangle \rangle _{a\widetilde{a}}.
\label{zero-mode-thermal-bst}
\end{eqnarray}
Here, the states from the various sectors of the thermal ground state are obtained by acting with the corresponding creation operators from $\OO (\beta_T)$ on to the thermal vacuum state, e. g. $|I\overline{J}(\b_T)\rangle\rangle = a^{I\dagger}_{1}(\b_T)\overline{a}^{J\dagger}_{1}(\b_T)|0(\b_T)\rangle\rangle$ etc. 

The thermal boundary state obtained in (\ref{thermal-boundary-state-final}) has an identical structure in terms of thermal string modes as the boundary state at zero temperature in terms of superstring modes, respectively. Therefore, it is appropriate to reefer to these states as to {\em thermal D-branes}. However, the thermal $D$-branes do not have a simple interpretation in terms of superstring modes at $T=0$ despite their coherent state form in terms of thermal string modes because the thermal string excitations represent a mixture of superstring an thermal reservoir excitations at $T =0$.

\section{Thermal boundary state entropy}

In this section we are going to derive the entropy of the thermal boundary states obtained in the previous section. According to the TFD formalism, the entropy operator of thermal string in $pp$-wave background can be factorized in to a bosonic $K^B$ and fermionic $K^F$ contributions, respectively, \cite{ubook,ng1}
\begin{equation}
K=K^{B}+K^{F}.
\end{equation}
The bosonic and fermionic terms can be written as sums over the type IIB GS superstring degrees of freedom as follows
\begin{eqnarray}
K^{B} &=&-\left\{ a_{0}^{\dagger }\cdot a_{0}\ln \left( \sinh^{2}\left(
\theta _{0}^{B}\right) \right) -a_{0}\cdot a_{0}^{\dagger }\ln \left( \cosh
^{2}\left( \theta _{0}^{B}\right) \right) \right\}
\nonumber
\\
&&-\sum_{n=1}\left\{ a_{n}^{\dagger }\cdot a_{n}\ln \left( \sinh^{2}\left(
\theta _{n}^{B}\right) \right) -a_{n}\cdot a_{n}^{\dagger }\ln \left( \cosh
^{2}\left( \theta _{n}^{B}\right) \right) \right\}
\nonumber
\\
&&-\sum_{n=1}\left\{ \overline{a}_{n}^{\dagger }\cdot \overline{a}_{n}\ln
\left( \sinh^{2}\left( \overline{\theta }_{n}^{B}\right) \right) -\overline{a}%
_{n}\cdot \overline{a}_{n}^{\dagger }\ln \left( \cosh ^{2}\left( \overline{%
\theta }_{n}^{B}\right) \right) \right\},
\label{bosonic-entropy}\\
K^{F} &=&-\left\{ S_{+}^{\dagger }\cdot S_{+}\ln \left( \sin^{2}\left( \theta
_{+}^{F}\right) \right) +S_{+}\cdot S_{+}^{\dagger }\ln \left( \cos
^{2}\left( \theta _{+}^{F}\right) \right) \right\}
\nonumber
\\
&&-\left\{ S_{-}^{\dagger }\cdot S_{-}\ln \left( \sin^{2}\left( \theta
_{-}^{F}\right) \right) +S_{-}\cdot S_{-}^{\dagger }\ln \left( \cos
^{2}\left( \theta _{-}^{F}\right) \right) \right\}
\nonumber
\\
&&-\sum_{n=1}\left\{ S_{n}^{\dagger }\cdot S_{n}\ln \left( \sin^{2}\left(
\theta _{n}^{F}\right) \right) +S_{n}\cdot S_{n}^{\dagger }\ln \left( \cos
^{2}\left( \theta _{n}^{F}\right) \right) \right\}
\nonumber
\\
&&-\sum_{n=1}\left\{ \overline{S}_{n}^{\dagger }\cdot \overline{S}_{n}\ln
\left( \sin^{2}\left( \overline{\theta }_{n}^{F}\right) \right) +\overline{S}%
_{n}\cdot \overline{S}_{n}^{\dagger }\ln \left( \cos ^{2}\left( \overline{%
\theta }_{n}^{F}\right) \right) \right\},
\label{fermionic-entropy}
\end{eqnarray}
where the dot stands for the sum over $I = 1,2,\ldots, 8$ or $a = 1,2,\ldots, 8$, respectively. The thermal $D$-brane entropy in $k_B$ units is given by the expectation value of the entropy operator in the thermal boundary state
\begin{equation}
S_{D}(\beta _{T})=\langle \left\langle B(\beta _{T})\right\vert K\left\vert
B(\beta _{T})\right\rangle \rangle .  
\label{thermal-D-brane-entropy}
\end{equation}%
In order to compute $S_{D}(\beta _{T})$, note that the thermal boundary
state (\ref{thermal-boundary-state-final}) can be written as 
\begin{equation}
\left\vert B(\beta _{T})\right\rangle \rangle ={\bf N}_{p+1}^{2}\sum \limits_{\mu =0}^{\infty }
\sum\limits_{\nu =0}^{\infty }\sum\limits_{\rho=0}^{\infty }
\frac{\left( -i\right) ^{\mu }}{\mu !\nu !\rho !}G^{\mu }\Sigma^{\nu }\widetilde{\Sigma }^{\rho }
\left\vert B_{0}\phi _{0}\right\rangle\rangle ,  
\label{expanded-thermal-D}
\end{equation}%
where $\left\vert B_{0}\phi _{0}\right\rangle \rangle $ represents the total string ground state contribution from (\ref{total-boundary-states-3}) and (\ref{total-boundary-states-4}). The combined action of the operators 
$G^{\mu}$, $\Sigma ^{\nu }$ and $\widetilde{\Sigma }^{\rho }$ creates bosonic and
fermionic superstring excitations in the directions $I=1,2,\ldots ,8$ and $a = 1,2,\ldots ,8$, respectively, in both left- and right-moving sectors. These excitations depend on the powers $\mu $, $\nu $ and $\rho $ in the expansion (\ref{expanded-thermal-D}). Let us denote the number of the corresponding excitations by $n_{B}^{I}(\mu ,\nu ,\rho ),\overline{n}_{B}^{J}(\mu ,\nu ,\rho ),m_{F}^{a}(\mu ,\nu ,\rho )$ and $\overline{m}_{F}^{b}(\mu ,\nu ,\rho )$, respectively. After some algebra, one obtains the following expressions for the bosonic entropy (\ref{thermal-D-brane-entropy})
\begin{eqnarray}
S_{B} &=&16\left( 2\sinh \pi m\right) ^{2\left( 3-p\right)
}\prod\limits_{k=1}^{\infty }\sum\limits_{l=1}^{\infty }\sum\limits_{\mu
,\nu ,\rho =0}^{\infty
}\sum\limits_{I,J=1}^{8}\sum\limits_{L,P=1}^{8}\sum\limits_{a,b=1}^{8}\sum%
\limits_{c,d=1}^{8}\frac{\left( -1\right) ^{\mu }}{\mu !\nu !\rho !} 
\nonumber\\
&&\left[M^{2\pi \left( \mu ,\nu ,\rho ,k\right) }\right] _{IJ}\left[
M_{(s)}^{2\omega \left( \mu ,\nu ,\rho ,k\right) }\right] _{ab}\left[ 
\widetilde{M}^{2\lambda \left( \mu ,\nu ,\rho ,k\right) }\right] _{LP}\left[
M_{(s)}^{2\chi \left( \mu ,\nu ,\rho ,k\right) }\right] _{cd},  
\nonumber \\
&&\times \left[ \beta _{T}\omega _{l}^{B}\delta (n^{\prime },n+1)+\ln \left(
1-e^{-\beta _{T}\omega _{l}^{B}}\right) \delta (n^{\prime },n)\right] ,
\label{bosonic-entropy-final}
\end{eqnarray}%
and the fermionic entropy 
\begin{eqnarray}
S_{F} &=&16\left( 2\sinh \pi m\right) ^{2\left( 3-p\right)
}\prod\limits_{k=1}^{\infty }\sum\limits_{l=1}^{\infty }\sum\limits_{\mu
,\nu ,\rho =0}^{\infty
}\sum\limits_{I,J=1}^{8}\sum\limits_{L,P=1}^{8}\sum\limits_{a,b=1}^{8}\sum%
\limits_{c,d=1}^{8}\frac{\left( -1\right) ^{\mu }}{\mu !\nu !\rho !}
\nonumber\\
&&\left[M^{2\pi \left( \mu ,\nu ,\rho ,k\right) }\right] _{IJ}\left[
M_{(s)}^{2\omega \left( \mu ,\nu ,\rho ,k\right) }\right] _{ab}\left[ 
\widetilde{M}^{2\lambda \left( \mu ,\nu ,\rho ,k\right) }\right] _{LP}\left[
M_{(s)}^{2\chi \left( \mu ,\nu ,\rho ,k\right) }\right] _{cd},  \nonumber \\
&&\times \left[ \beta _{T}\omega _{l}^{F}\delta (m^{\prime },m+1)+\ln \left(
1+e^{-\beta _{T}\omega _{l}^{F}}\right) \delta (m^{\prime },m)\right] ,
\label{fermionic-entropy-final}
\end{eqnarray}%
respectively. Here, $\pi \left( \mu ,\nu ,\rho ,k\right) ,\omega \left( \mu
,\nu ,\rho ,k\right) ,\lambda \left( \mu ,\nu ,\rho ,k\right) ,\chi \left(
\mu ,\nu ,\rho ,k\right) $ represent powers at which the corresponding
matrices contribute to the entropy and $k$ and $l$ label the superstring
modes. For the sake of simplicity, the sums over $\pi ,\omega ,\lambda $ and $\chi $
were not written explicitely and it should be understood that these sums
should be taken together with the sums over $\mu ,\nu $ and $\rho $. Also,
we have used the short-hand notation $n^{\prime }=n_{B}^{\prime I}(\mu ,\nu
,\rho ),m^{\prime }=m_{F}^{\prime a}(\mu ,\nu ,\rho )$, etc. Thus, the delta
functions $\delta (n^{\prime },n)$ are short-hand notation for products of
Kronecker symbols, e. g. $\delta (n^{\prime },n)=\delta _{\mu ^{\prime }\mu
}\delta _{\nu ^{\prime }\nu }\delta _{\rho ^{\prime }\rho }\delta
_{k^{\prime }k}\delta _{I^{\prime }I}\delta _{kl}\delta _{IL}\delta
_{n^{\prime }n}$, where the last Kronecker symbol is for $n^{\prime }$ and $n
$ viewed as natural numbers. The above relations can be used to calculate
the entropy of zero modes. To this end, one should take for the bosonic
entropy $l=0$, drop the sum over $l$, and divide the r. h. s. of (\ref%
{bosonic-entropy-final}) by 2 since for zero modes there is no left- and
right-moving contribution in $S_{B}$. The entropy of either $+$ or $-$
fermionic zero modes can be obtained from (\ref{fermionic-entropy-final}) by
droppying the sum over $l$, dividing by 2 for any of $\pm $ modes and
replacing $\omega _{l}^{F}$ by $\omega _{\pm }^{F}$. Note that $\left[
M^{2\pi }\right] _{IJ}=\left[ \widetilde{M}^{2\lambda }\right] _{IJ}=\delta
_{IJ}$ since $M_{IJ}= \mbox{diag} \left( \pm ,\pm ,\ldots ,\pm \right) $. Therefore, the
sums over $I,J$ and $L,P$ give a contribution of 16 that multiplies the front factor for $l\neq 0$.

\section{Conclussions}

To conclude, in this paper we have constructed the thermal boundary states for the type IIB GS superstring in 
{\em pp}-wave background and we have determined their entropy. 

The thermal $D$-branes were obtained by thermalizing the superstring and its boundary conditions as in the flat spacetime \cite{ivv12}. The main subtlety here is that in the definition of the total lagrangian given in (\ref{lagrangian-density}), the relation (\ref{tilde-lagrangian}) should be chosen rather than the tilde conjugation. Indeed, if $\tilde{{\cal L}}(\tilde{\Phi})=[{\cal L}(\Phi )]{\tilde{~}}$ had been chosen instead of $\tilde{{\cal L}}(\tilde{\Phi})={\cal L}(\tilde{\Phi})$, the tilde superstring would have had different supersymmetry charges as discussed in Section 3.1. The existence of the boundary states that satisfy the
boundary conditions with these charges remains as an open problem. Beside that, the energies of the fermionic modes from (\ref{total-hamiltonian-1}) and (\ref{total-hamiltonian-2}) would add up, which means that the interaction between
the superstring and the reservoir is not thermal in the fermionic sector. The thermal boundary state (\ref{thermal-boundary-state-final})  factorizes in a thermal non tilde and a thermal tilde contribution, respectively. However, despite its similarity with the total boundary state for the superstring and the thermal reservoir at zero temperature, its interpretation in terms of zero temperature superstring modes should be different since thermal $D$-brane states cannont be factorized in any simple way in terms of superstring excitations at $T = 0$. These conclussions are similar to those obtained for the thermal boundary states in the flat spacetime \cite{ivv12}. The thermal boundary state entropy was computed using the entropy operator of the superstring at zero temperature as in TFD (see Appendix A.) The thermal $D$-brane entropy is different from that of the thermal string which simply sums the contribution of the superstring oscillators \cite{ng1} because the thermal boundary state is not the vacuum state of the thermal string. An unsolved important problem is to find a formulation of the thermal $D$-brane in which the thermal boundary state could be interpreted as a ground state.

{\bf Acknowledgments}

I would like to acknowledge to J. A. Helay\"{e}l-Neto, S. A. Dias and A. M. O. de Almeida for hospitality at 
LAFEX-CBPF where part of this work was done.

\appendix

\section{Review of the TFD Formalism}

In this appendix we are going to briefly review the TFD formalism. The first step in the TFD method is to duplicate the operators corresponding to the degrees of freedom of the system under consideration. The second step consists in defining the properties of the thermal ground state. 

The system degrees of freedom are doubled by associating to any operator $A$ an identical operator $\tilde{A}$. The two sets 
$\{ A \}$ and $\{ \tilde{A} \}$ are considered independent, i. e. 
\be
[A,\tilde{B}]_{\pm}=0,
\label{indeptildeops}
\ee
for any $A$ from $\{A\}$ and any $\tilde{B}$ from $\{\tilde{A}\}$. Here, $+$ denotes the anti-commutator taken when both $A$ and $\tilde{B}$ are fermionic. Otherwise, the commutator should be taken. The two sets of operators $A$ and $\tilde{A}$ are in one-to-one correspondence through the tilde-conjugation rule given by the following axioms
\bea
(A_1 A_2 )\ \tilde{} &=& \tilde{A}_1\tilde{A}_2,\label{tilde1}\\
(c_1 A_1 + c_2 A_2 )\ \tilde{} &=& c^{*}_1 \tilde{A}_1 + c^{*}_2 \tilde{A}_2, \label{tilde2}\\ 
(A^{\dagger})\ \tilde{} &=& (\tilde{A})^{\dagger},\label{tilde3}\\
(\tilde{A})\ \tilde{} &=& A,\label{tilde4}
\eea
for all complex numbers $c$, $c_i$ and all operators $A$, $A_i$. Here, $c^*$ denotes the complex conjugate of the number $c$. The tilde-conjugation is an involution on the algebra of the total string operators.

The thermal ground state is specified by the thermal state condition. In general, this is defined by taking asymmetric bra and ket ground states that satisfy the following relations
\bea
\langle 1 | &=& \sigma^{*}\langle 1 | \tilde{A}^{\dagger},\label{thermalbra}\\
A|0(\b_T)\rangle &=& \sigma e^{\b_T \hat{H}}\tilde{A}^{\dagger}e^{-\b_T \hat{H}}|0(\b_T) \rangle,\label{thermalket}
\eea
for any $A$, where $\sigma = +1$ for bosons and $\sigma = i$ for fermions and 
\be
\hat{H} = H - \tilde{H}.
\label{totalHamiltonian}
\ee
The phases of the thermal ground states (thermal vacua) are not unique and it can be chosen such that
\be
(\langle 1|)\ \tilde{} = \langle 1 | ~~,~~(|0(\b_T) \rangle\ \tilde{} = |0(\b_T)\rangle.
\label{tildevacua}
\ee
These relations can be regarded as an operator representation of the KMS condition. The justification for taking these relations as axioms of the TFD method is that the vacuum expectation value between the above vacua corresponds to the thermal average as follows. By expressing $| 1 \rangle $ and $ |0(\b_T) \rangle $ as 
\be
|1 \rangle = \sum_{n}|n\tilde{n}\rangle~~,~~|0(\b_T)\rangle = \frac{e^{-\b_T H}|1\rangle}{\mbox{Tr}[e^{-\b_T H}]},
\label{explicitthermalvacua}
\ee
where $|n\tilde{n}\rangle$ are the orthonormal basis vectors in the total Hilbert space, one obtains
\be
\langle 1 | A | 0(\b_T) \rangle = \frac{\mbox{Tr}[A \ e^{-\b_T H}]}{\mbox{Tr}[e^{-\b_T H}]}.
\label{fundamentalTFD}
\ee
One can note that $| 1 \rangle $ does not depend on the choice of the basis vectors and that it is a kind of identity state (hence the notation). The information about the thermal ground state is contained in the ket ground state 
$| 0(\b_T) \rangle$. The evolution of the thermal system is generated by the total hamiltonian $\hat{H}$ which can be obtained from a total lagrangian
\be
\hat{\LL} = \LL - \tilde{\LL}.
\label{totalLagrangian}
\ee 
In the canonical quantization, the formalism takes a simple operatorial form. For a linear oscillator the following relations hold
\bea
|1\rangle_B = e^{a^{\dagger}\tilde{a}^{\dagger}}|0\rangle_B,\label{bosonunit}
|1\rangle_F = (1 + i \ a^{\dagger}\tilde{a}^{\dagger})|0\rangle_F,\label{fermionunit}
\eea
for the bosonic and fermionic oscillators, respectively. Thus, the thermal bra and ket ground states can be taken symmetric and the formalism is the same as the formalism at zero temperature. The mapping from zero temperature to finite temperature is generated by the Bogoliubov operator
\be
G = - i\theta(\b_T)(a\tilde{a} - \tilde{a}^{\dagger}a^{\dagger}),
\label{Bogoliubovoscillator}
\ee
which is conserved
\be
[\hat{H},G] = 0.
\ee
In general, the Hilbert space at zero temperature and the thermal Hilbert space are not isomorphic for systems with an infinite number of degrees of freedom (see for further details \cite{ubook}.)

The entropy of a quantum field in terms of creation and annihilation operators in $k_B$ units is given by the expectation value of one of the following operators in the thermal vacuum state \cite{ubook} 
\bea
K^B &=& - \sum_{n=1}^{\infty}\left(a^{\dagger}_n a_n\, \mbox{log}\,\mbox{sinh}^2\,\theta^{B}_n - a_n a^{\dagger}_n\, \mbox{log}\,\mbox{cosh}^2\,\theta^{B}_n \right),\label{Kb1}\\
K^F &=& - \sum_{n=1}^{\infty}\left(a^{\dagger}_n a_n\, \mbox{log}\,\mbox{sin}^2\,\theta^{F}_n + a_n a^{\dagger}_n\, \mbox{log}\,\mbox{cos}^2\,\theta^{F}_n \right),\label{Kf1}
\eea
where $K^B$ and $K^F$ stand for the entropy of the bosonic and fermionic field, respectively.
The form of $\theta$'s is given by the following relations \cite{ubook}
\be
\theta^{B}_{n}(\beta_T ) = \mbox{arccosh} (1-e^{-\beta_T \omega^{B}_n})^{-\frac{1}{2}}~,~
\theta^{F}_{n}(\beta_T ) = \mbox{arccos} (1+e^{-\beta_T \omega^{F}_n})^{-\frac{1}{2}}.
\label{thetas}
\ee
In the case of type IIB GS superstring in {\em pp}-wave background, by including in the total hamiltonian the level matching conditions \cite{ng1}, the $\theta$'s can be modified to the expressions given in (\ref{theta-zero-modes})-(\ref{theta-fermions}). The entropy operator takes into account only superstring oscillators at $T =0 $. Including operators from the tilde superstring would mean to take the average over the reservoire degrees of freedom, too.

\end{document}